# Thinging-Based Conceptual Modeling: Case Study of a Tendering System

**Sabah Al-Fedaghi and Esraa Haidar**

*Department of Computer Engineering, Kuwait University, Kuwait City, Kuwait*

**Abstract:** In computer science, models are made explicit to provide formality and a precise understanding of small, contingent "universes" (e.g., an organization), as constructed from stakeholder requirements. Conceptual modeling is a fundamental discipline in this context whose main concerns are identifying, analyzing and describing the critical concepts of a universe of discourse. In the information systems field, one of the reasons why projects fail is an inability to capture requirements in a way that can be technically used to configure a system. This problem of requirements specification is considered to have "deficiencies in theory". We apply a recently developed model called the Thinging Machine (TM) model which uniformly integrates static and dynamic modeling features to this problem of requirements specification. The object-Oriented (OO) approach to modeling, as applied in Unified Modeling Language, is by far the most applied and accepted standard in software engineering; nevertheless, new notions in the field may enhance and facilitate a supplementary understanding of the OO model itself. We aim to contribute to the field of conceptual modeling by introducing the TM model's philosophical foundation of requirements analysis. The TM model has only five generic processes of things (e.g., objects), in which genericity indicates generality, as in the generic Aristotelian concepts based on abstraction. We show the TM model's viability by applying it to a real business system.

## Introduction

Modeling is used to understand and shape the world and is a foundational technique, in that "There is hardly a domain of inquiry without models" (Frigg and Nguyen, 2017), e.g., the solar system as well as atoms, cells and electricity. As a representation of the selected part or aspect of the world being investigated, a model can explain the nature of its subject matter (Frigg and Nguyen, 2017). Although modeling has been employed for ages in virtually all disciplines, the form of models has been fairly recently made explicit in computer science, where it is utilized to "provide formality and a precise understanding of what is a well-formed model to the communication between humans and machines" (Hölldobler *et al.*, 2017).

We focus on conceptual modeling, specifically on the Object-Oriented (OO) approach as applied in the Unified Modeling Language (UML, with special attention to the notion of class/object). OO modeling is by far the most applied and accepted standard in software engineering. Nevertheless, new developments in the field may enhance and facilitate a supplementary understanding of the OO model itself. Our main concern is with deficiencies in requirement analysis theory within software engineering. This eventually leads us to our main goal of proposing a new conceptual modeling technique with a single construct, called *thimac* (thing/machine), which unifies the static and dynamic features of things (e.g., objects). We show the viability of the thimac notion by applying it to a real business system.

### Conceptual Modeling

According to Guizzardi and Halpin (2008), "Conceptual modeling (CM-including information or data modeling) is a fundamental discipline to several communities in computer science. Its main objective is concerned with identifying, analyzing and describing the essential concepts and constraints of a universe of discourse with the help of a (diagrammatic) modeling language that is based on a set of basic modeling concepts." CM has an enormous impact on information system artifacts because conceptual models determine the acceptability and usability of the product to be built (Lauesen and Vinter, 2001). It is the most important part of requirements engineering and the first phase toward designing an information system (Hossain and Schwitter, 2018). In this context, the focus is on small, contingent "universes" constructed from stakeholder requirements (Singh, 2011). This application of modeling suggests how key ideas from the philosophy may be fruitfully adapted and thereby help to improve research and practice (Singh, 2011). While the debate on philosophy may or may not be seen as essential, engagement in philosophy cannot be avoided since a "good part of the answer to the question 'why philosophy?' is that the alternative to philosophy is not no philosophy but bad philosophy" (Recker, 2005).

CM as a theoretical enterprise has underlying philosophical schemes. In our case, we focus on the problem of *representation*, which deals with the problem of ontology: Kinds of objects in a model, including their static and dynamic features. Here, we view ontology modeling as a form of CM. Ontology concerns the kinds of objects and constructs that are sufficient for describing reality. For example, the BWW (Bunge-Wand-Weber) ontology (Bunge, 1977) has been applied in conceptual modeling as a reference point in specifying reality constructs (Wand and Weber, 2002). In this framework, "The universe of discourse comprises immutable objects and object structures that exist as empirical entities. A conceptual model is, in this perception, understood as an objective perspective through which observers can perceive unbiased reality" (Recker, 2005).

CM produces a technology-independent specification that precisely describes the domain entities for communication, learning and problem-solving. This conceptual specification is transformed into a logical design specification by considering a number of design issues (Guizzardi and Halpin, 2008). A conceptual model is a medium with which to foster communication with prospective users and provides a basis for system implementation (Frank, 1999). It is a commonly accepted approach to overcoming communication problems (Wand and Weber, 1993). Furthermore, conceptual models help analysts understand a domain, provide input into the design process and document the requirements (Recker, 2005). Examples of languages in this context include UML and Entity-Relationship (ER) notation (Chen, 1976).

Accordingly, as described by Recker (2005), there has been a "flooding" of CM approaches. "The area of CM is, however, coined by a juxtaposition of different terms and concepts" (Recker, 2005).

*Problem: Deficiencies in Theory of Requirements*

In the information systems area, one of the reasons why projects fail is miscommunication leading to an inability to explicitly specifying the requirements in a way that can be technically used to configure a system such as a process model (Ribbert *et al*., 2004). This problem of requirements specification is considered to have "deficiencies in theory" (see sources in Ribbert *et al*., 2004). Difficulties related to work in this area are centered on several wide issues as described by Ribbert *et al*. (2004), which include:

- How to model the "existence of a real world" in terms of ontology of "what is" and "how it is"; and
- Issues concerning the relationship between objects and subjects, in terms of whether things in the real world can in principle be objectively recognized (correctly)

The problem of deficiencies in theory partially results from the fact that English is typically utilized as a means of identifying concepts (e.g., classes or objects) and building a model. Models are developed by experts who need to be members of a language community (Ribbert *et al*., 2004). However, the model serves as a basis for non-experts within the system domain to understand the different facets of the domain. Thus, it is suggested that models representing experts' knowledge need to be provided to users and developers as non-experts. The models must abstract from certain aspects, such as technical (e.g., software) or organizational details. In discussing this issue, the following observations are modifications of ideas in Ribbert *et al*. (2004):

- Models must be expressed in a unified language that can be understood by the targeted users (participants in internal processes or developers), suitable from an expert's perspective and usable from a user's perspective
- Models must have high abstraction to represent all aspects of the entire system
- Complexity is reduced by providing multiple levels of sophistication in descriptions for the same model

*Philosophical Modeling Approach to Solutions*

One of the foundational philosophical schemes of OO modeling is the Greek philosophy of form and matter. The notions of OO modeling have been abstracted into key ingredients of systems analysis with the key notion of *object*. Objects, in software, are viewed just as nuts, bolts and beams are in construction design (de Champeaux *et al*., 1993). According to Grässle *et al*. (2005), the basis of the OO approach is "as good as possible" of a representation of something that exists in the real world. A fundamental notion in this context is the *class*. A class defines an object's properties and methods, called an *instance* of the class. A class is a Platonic notion and a template for objects in which forms serve as patterns for real-world things.

UML has emerged as a language for conceptual modeling, specifically for "communicating between users and developers in understanding and eliciting requirements and also for documenting the outcome of analysis" (Lu and Parsons, 2005). The class diagram is the most fundamental UML diagram (Szlenk, 2006) and is "a central modeling technique that runs through nearly all object-oriented methods" (Stotts, 2007). Class diagrams provide an overview of systems and are utilized for purposes such as understanding requirements and describing the target system's design in detail. It is the best-known view of the OO approach and often the only diagram that is constructed (Grässle *et al*., 2005). The class diagram is useful throughout the entire software development process, from early domain analysis stages to later maintenance stages (Washizaki *et al*., 2010). A great deal of research on OO design has explored how to identify classes and their relations and class diagram layouts have been examined from different perspectives, such as visibility, juxtaposability and aesthetics (Eichelberger, 2003).

To identify classes, Osis and Asnina (2010) developed a graph transformation from "topological functioning modeling" to a conceptual class to enable "the definition between domain concepts and their relations to be established." Stotts (2007) claims that "the lines between the [conceptual, specification and implementation] perspectives [when using use class diagrams] are not sharp and most modelers do not take care to get their perspective sorted out when they are drawing." Generally, "the biggest danger with class diagrams is that you can get bogged down in implementation details far too early." To combat this, the conceptual perspective is adopted (Stotts, 2007). According to a university document (DCS, 2010),

It is only fair to point out that not all experts support the UML effort and it comes under regular and harsh criticism, some of it fair. For example, one criticism is that there is not good enough *integration between the different components* of the UML (e.g., between use case and class modeling). (Italics added.)

We apply a recently developed model called the Thinging Machine (TM) model, which uniformly integrates static and dynamic features, to the theoretically deficient problem of requirements specification. We aim to contribute to the philosophical basis of conceptual modeling by providing the ontological foundation of the TM model. This model is unique in terms of the following:

1. It incorporates a complete ontological unity between things (e.g., objects) and processes (called machines). The detail of this unity is defined through the intrinsic structure, in terms of a network of what are called thimacs (*thi*ng/*ma*chine), which provide an alternative conceptualization to classes and subclasses
2. It is built upon five generic operations that are applied to things (e.g., objects)
3. It integrates a system's static and dynamic features by superimposing events (and hence time) over the same diagrammatic static representation to specify the system's behavior

The next section provides a more elaborate description of the TM model, which has been applied to several real systems, such as phone communication (Al-Fedaghi and Aldamkhi 2019), physical security (Al-Fedaghi and Alsumait, 2019), vehicle tracking (Al-Fedaghi and Al-Fadhli, 2019), computational thinking (Al-Fedaghi and Alkhaldi, 2019) and information leakage (Al-Fedaghi and Behnehani, 2018).

To illustrate TM modeling and provide a contrasting instance to the OO approach, Sections III and IV apply TM to identifying UML classes. To demonstrate modeling in TM, Section V presents an actual government organization. More information about the TM model can be found in Al-Fedaghi (2019a; 2019b; 2019c).

## Introduction to Thinging Machines

We start our discussion of the TM model with the notion of *things*, which originated with the German philosopher Heidegger (1975). According to Heidegger (1975), a thing is self-sustained, self-supporting or independent-something that stands on its own. For example, a *tree* is a thing through which sunlight, water, carbon dioxide, minerals in the soil and so on flow. Through a series of processes, the tree-thing transforms those flows of matter into various sorts of cells (Bryant, 2012). Heidegger (1975) encourages further research on "generic processes" applied to things. We now focus on five of these processes and claim that they are sufficient for modeling purposes.

### Notion of "Thing" in the TM Model

We postulate that only five generic processes of things exist: Things can be *created*, *processed*, *released*, *transferred* and *received*. For instance, suppose that $t$ is a thing. To describe the generic processes that can be applied to $t$ in a given *system, S* (whose definition will be discussed later), the following argument presents an informal justification for these five processes:

- Thing *t* either comes from outside of *S* (**transferred in**) or is internally generated (**created**)
- When *t* is transferred from outside of *S*, it is either rejected or **received** as one of the system's things.
- Thing *t* in *S* may be **transferred outside** of *S*
- The thing may be put into the **released** state until a channel is open for transferring it outside
- During its residency in *S*, *t* may be **processed** (changed)

These five generic processes form an abstract machine called a TM. The TM approach is most meaningfully communicated in a diagrammatic way, as shown in Fig. 1, where the elementary processes are called the *stages* of a TM. The TM in Fig. 1 is a type of abstract *machine* that handles things. The flow (solid arrows in Fig. 1) among the five stages signifies *conceptual* movement from one machine to another or among the stages of a machine. The TM stages can be described as follows:

**Arrived**: A thing reaches a new machine.
**Accepted**: A thing is permitted to enter the machine. If arriving things are always accepted, then *arrive* and *accept* can be combined into the *received* stage. For simplification, the examples in this paper assume the received stage.
**Processed** (changed): A thing undergoes some kind of transformation that changes it without creating a new thing.
**Released**: A thing is marked as ready to be transferred outside of the machine.
**Transferred**: A thing is transported to somewhere outside of the machine, or from somewhere outside of the machine.
**Created**: A new thing is born (created) within a machine. This is the starting point of a thing in a system. The term *create* comes from creativity with respect to a system (i.e., constructed things from already created things, or emergent things that appear in a system from somewhere). In the TM model, creation encompasses existence.

Additionally, the TM model includes **memory** and **triggering** relations (represented as dashed arrows) among the processes' stages (machines), as illustrated later.

The genericity of processes indicates generality as in the generic Aristotelian concepts based on abstraction. TM classifies processes into five types that are applied to all entities that have common subject-oriented and OO aspects, as will be clarified later. Genericity implies that a generic process cannot be reduced to the other four generic processes. *Creating* a new thing cannot be the result of changing (processing) an old thing. No matter how a thing is released, no new thing is produced. Transferring does not reform a thing into a new thing and receiving a thing implies that it was created previously.

*Thimac: A Thing is a Machine and a Machine is a Thing*

A TM thing is defined as what can be created, processed, released, transferred and/or received. Simultaneously, in this modeling approach, a thing is also a five-dimensional structure referred to as an (abstract) machine. From a different perspective, *machines are things* that are "operated on"; that is, they are created, processed, released, transferred and received. Machines are intertwined with the world through the inseparable coherence integrated in these two poles of an entity's being: *Being a thing* that flows through machines and *being a machine* that handles things. According to our thesis, these are equally irreducible modes of being.

Therefore, we can view the Heideggerian tree as a thimac (a word formed from the first three letters in *thi*ng and *mac*hine) through its network of subthimacs: Flow of sunlight, water, carbon dioxide, minerals in the soil, etc. Through the five generic stages, the machine transforms those flows of matter—the other machines that pass through it—into various sorts of cells (Bryant, 2012). This tree exits without multiplicity, regionicity (actual space) or other so-called secondary categories. Additionally, the TM description of the tree incorporates time, to dress the *static* model with instances of events (examples will be given later).

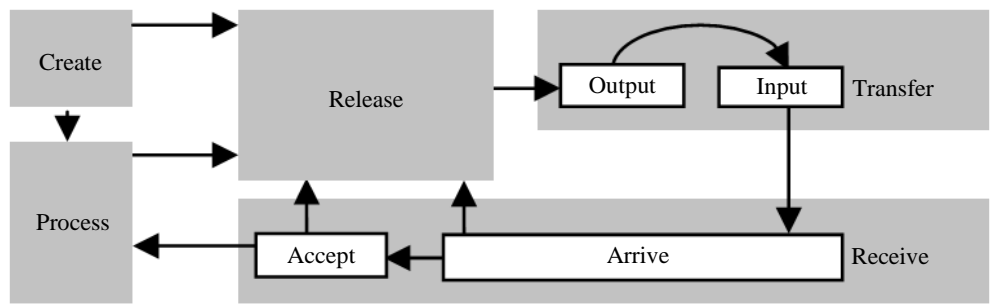

**Fig. 1:** Thinging machine

The TM model is used for modeling specific systems, not for representing "reality" per sé. The example of the tree is discussed to emphasize that some subthimacs are relevant in the undergoing modeling and that the grand thimac (system; e.g., the Heideggerian tree) combines all subthimacs without excluding any of them. The entire TM diagram is a grand thimac that forms an architectural whole or totality. The static TM description is where things are projected into "conceptual being". A TM model's organization remains invariant, while the dynamic TM model is constantly transforming. Organization consists of an assembly of subthimacs related by flows and triggering. The flow reflects input/output interactions and triggering is a non-input/non-output contact.

The system's unity (system/grand thimac) is maintained by the TM structure of flow, the five generic stages and triggering. It is supplied by multiple flows of things that are created in many thimacs. This implies unity through intrinsic structure, with the possibility of multiple so-called substances. In the information system context, the grand thimac interweaves within it all users and other internal/external supplies (creators) of things, data, information, actuators, signals, images, etc.

*Thimacs in a System*

Thimacs, as a founding category of being, replace traditional categorizations, properties and behaviors. They determine what an entity is as a thing and as a machine. *Thimacing* is a conceptualization of a thimac network used to express abstractions of the state of affairs in a given portion of reality. A TM model is the abstraction of a sphere articulated according to a domain conceptualization. Note that the thimac notion is not new.

In physics, entities at the subatomic level must be regarded as both particles and waves to enable a full description and explanation of observed phenomena (Steiner, 1985). According to Sfard (1991), abstract notions can be conceived in two fundamentally different ways: Structurally, as objects/things (static constructs) and operationally, as processes. This paper adopts this notion of duality in conceptual modeling, generalizing it beyond mathematics and its utilization in software engineering modeling. Structural conception means seeing a notion as an entity with a recognizable static structure. The operational way of thinking emphasizes the dynamic process of performing actions. A model is a description of a given domain independent of technological choices that could impact the implementation of a system based on itself.

*Example*

Flow indicates a change to a TM's spatial form (different stage or machine). A TM flow encompasses the classical notion of motion; thus, heat flowing to water triggers an increase in the water temperature (. 2 and 3). In Fig. 2, heat (a thing) is created (it appears), then is released and transferred. The water (as a thimac) receives and processes the heat, which triggers (dashed arrow) the creation of an increase in temperature, which is processed (takes its course).

Figure 3 shows the dynamic features, which are supimposed (will be further defined later) on the diagram of Fig. 2. Two events are recognized: The flow of heat (yellow) and the increase in temperature (orange). Figure 4 shows the system's behavior in terms of the two events: More heat results in a greater increase in temperature. We can see here the "nature" of the static TM (Figure 2) description, in which the arrows in the static diagram represent a map of dry rivers (red and purple arrows).

Philosophically, the static TM model forms the basis upon which potentialities that are materialized through events are modeled. It has the capacity to be real without being actual (DeLanda, 2015). Potentiality and actuality are Aristotelian notions that refer to movement from the possible to the real. They are related to a TM's passage from static description to dynamic specification by applying events (and time) over the original TM diagram. The TM's dynamic specification involves multiplicity, e.g., looping (Aristotelian number), the region of the event (Aristotelian space) and the flow of time (approximately, Aristotelian motion).

While a thimac reflects the idea of unity, the details of this unity are defined through the system's intrinsic structure in terms of its thimacs and its network of subthimacs. These replace the typical conceptualization of classes and subclasses. The actuality (the dynamic system) is related to the idea of truth (i.e., data and events reflect the reality of the system and its being).

## Applying the TM Model to Identifying Classes

As a further illustration of the TM modeling approach, we apply it to a known problem in the field. Finding classes, as the first step in capturing requirements, is a central decision in OO software systems; making such decisions correctly takes talent, experience and luck (Meyer, 1997). In function-oriented design, we would concentrate on the verbs; in OO design, we underline the nouns, which describe objects (Meyer, 1997). In the TM approach, we search for processes (TM machines) that involve creating, processing, releasing, transferring and/or receiving things.

*UML-Based Methods of Identifying Classes*

Consider a sample approach to identifying classes called noun extraction. de Champeaux *et al*. (1993) utilized this method of identifying classes in the context of an example of a Bank (B), which is described as follows (classes shown in italics).

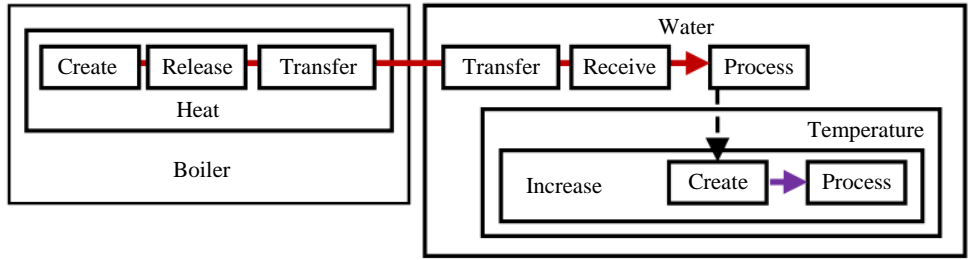

**Fig. 2:** The TM model of heating water

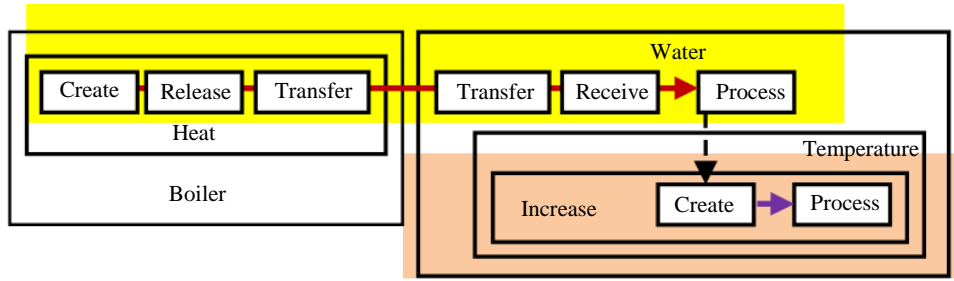

**Fig. 3:** Events in the TM model of heating water

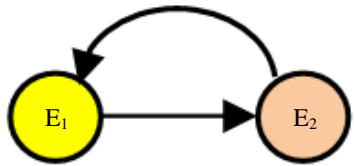

**Fig. 4:** Repeating the events

Every *branch office* has equipment to maintain the *accounts* of its *clients*. All *equipment* is networked together. Each *ATM* is associated and connected with the *equipment* of a particular branch office. *Clients* can have *checking, savings* and *line of credit accounts*, all conveniently interconnected… (de Champeaux *et al*., 1993). de Champeaux *et al*. (1993) select the noun phrases with the classes of branch office, account, client, equipment, ATM, etc.

In *event-based* class identification, according to Singh *et al*. (2010), a large number of diagrams need to be analyzed before arriving at a final class diagram. Singh *et al*. (2010) give an example of a list of events from an online reservation system that includes:

- A customer views the tour information (external event)
- A customer makes a reservation while on a tour (external event)
- A customer cancels a reservation while on a tour (external event)

Then, Singh *et al*. (2010) identified events that are explicitly specified in the requirements statements or added by a domain expert. The events lead to a class diagram specification being derived. In OO methodology, an event is generally an external stimulus from one object to another that occurs at a particular point in time. It is a transmission of information from one object to another. A scenario is a sequence of events that occurs during one particular execution of a system (Nath, 2020).

Alternatively, when applying the initial TM-based thinking of this problem, we can search for machines. As a result, in the example of an online reservation system, a static TM model is produced, as shown in Fig. 5.

We isolate and cut the problem space up into TM machines that handle things that flow: Machines that handle requests; machines that provide lists of offers; and reservation machines, for which things are created, processed, released, transferred and received in each case.

Figure 6 shows some of the TM events (dotted circles) in part of Fig. 5. The notion of what an event is in software modeling still seems unsettled. For example, the OMG's (2000) UML specification defines an event as a noteworthy occurrence; in Rumbaugh *et al*.'s (1991) words, "an event is a noteworthy change in state." In TM, we can identify events from the static TM description (e.g., Fig. 5) based on elementary events that correspond to the five generic processes. Events at many levels are constructed from lower-level events. Humans seem to focus on events in the middle of event hierarchies. The same phenomenon is applied to classes in the OO methodology. According to Taivalsaari (1997), in "class hierarchies in object-oriented programming, the basic classes typically end up in the middle of the class hierarchy. In contrast, those classes that are at the top (root) or at the bottom (leaves) of the hierarchies are typically of less interest, either because they are overly generic or overly specific for the purpose of examination."

Further description of the notion of events in the TM model will be provided during the following discussion of the examples.

*How to Represent a Class in the TM Model*

The TM model can enhance different notions within the OO methodology. Consider the notion of class. The term *class* has two somewhat different meanings:

- It is the pattern according to which objects are created
- It is the set of objects that have been created according to that class

The class, as a pattern, dictates the characteristics and behavior of objects that are created from it. Authors of OO books, including Grässle *et al.* (2005) and Weisfeld (2009), like to compare a class to a cookie cutter, which can be used to cut cookies (objects of the class) from dough. In the example, the dough is shown in Fig. 7.

In the TM model, an entity is not a pattern and an object, but an integration of a thing and a machine (*thimac*). A thing, in this vocabulary, refers to a family of instances of things, just as class and objects do. These instances flow in thimacs that reflect a particular gestalt. As shown in Fig. 8, the dough is a machine (circle 1) with submachines and itself is a thing that can be created, released and transferred (2). It has a circular pattern. The cutter (3) processes (4) the dough to trigger (5) the creation of cookies (6). The cookies have a star pattern (7) and form a collection (8).

Additionally, the TM embeds the dough/cookie system's behavior in terms of events. An event in TM is a thimac. For example, Fig. 9 shows the event *The dough has been processed by the cutter*. The *region* of the event is where the event occurs. For simplicity, we will represent events only by their regions. Accordingly, we select the following events, as shown in Fig. 10:

- Event 1 ($E_1$): *The dough is created in a circular shape*
- Event 2 ($E_2$): *The dough is processed by the cutter*
- Event 3 ($E_3$): *Cookies are created in a star shape*

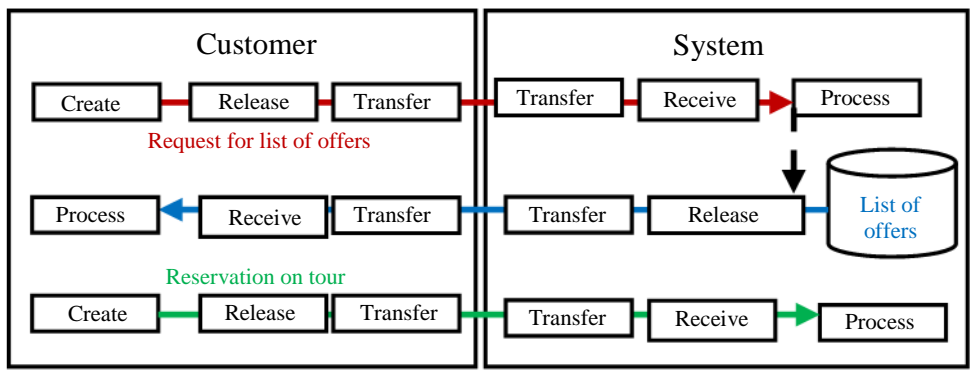

**Fig. 5:** TM model of Singh *et al.*'s (2010) event *Customer views tour information*

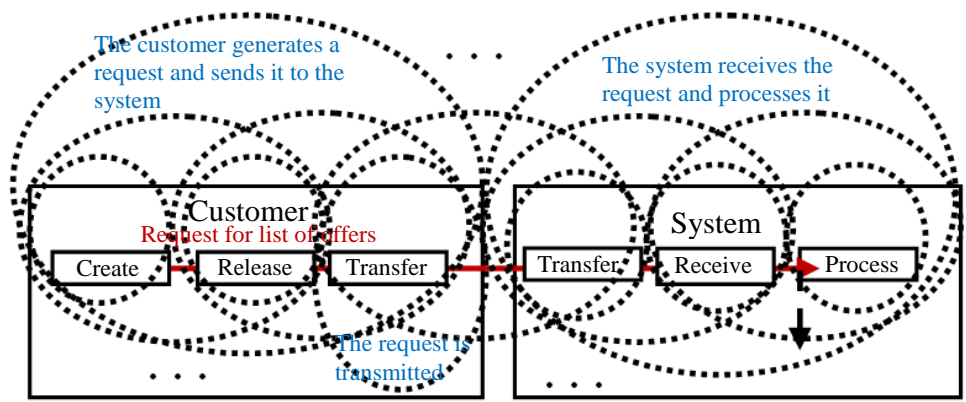

**Fig. 6:** Possible events in the TM model

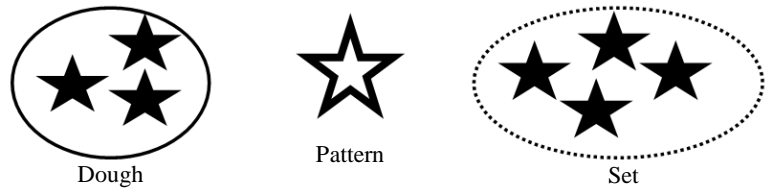

**Fig. 7:** Illustration of the class in the cookie-cutter example (redrawn from Grässle *et al*., 2005)

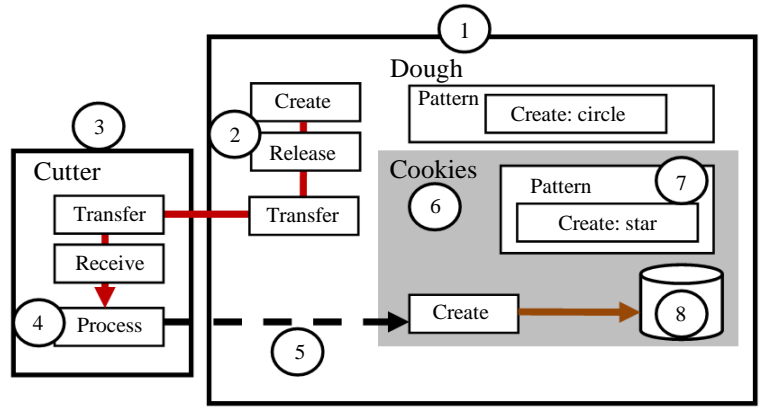

**Fig. 8:** The dough as a thimac

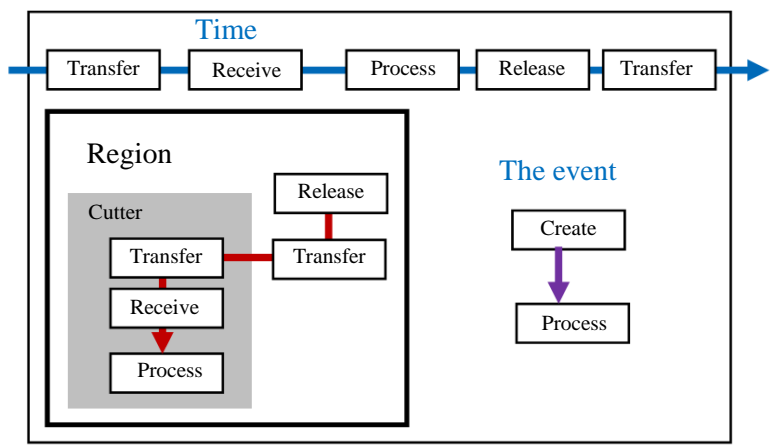

**Fig. 9:** The event *the dough has been processed by the cutter*

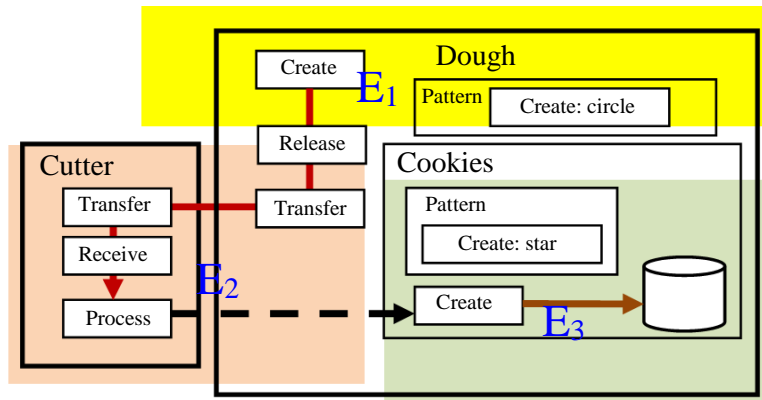

**Fig. 10:** The events in the dough/cookie example

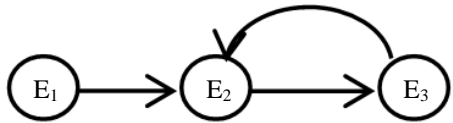

**Fig. 11:** Behavior of the dough/cookie system

Figure 11 shows the system's behavior as a chronology of events.

## Applying the TM Model to a Case Study: Tendering System

As mentioned previously, in the information systems field, one of the reasons why projects fail is the inability to capture requirements in a way that can be technically used to configure a system. In this section, we introduce an actual case study that involves capturing requirements. We now provide a sample of this problem in terms of designing a tendering system in a real organization (the second author's workplace) applying the TM model. The case involved in this paper is a tendering system that describes the actual process of how a vendor can register itself in the system in order to apply its purchase orders (POs), which can be described as follows:

(1) A vendor acquires an account
(2) The account must be activated
(3) A registration fee is paid
(4) The vendor account is activated
(5) The vendor fills out the purchase order application

Figure 12 and 13 show sample representations of the current documentation of the tendering application using UML diagrams.

Tendering processes are complex and involve many business procedures, such as tender specification preparation, tender awarding and contract monitoring. A tendering system often needs to communicate with other systems, such as supply, order and purchase systems, to complete its procedures (Ng *et al.*, 2007). In a traditional paper-based bidding process, after a tender is released, suppliers must provide quotations to the tendering system so that they can be ranked by certain tender requirements before the tender contract is selected and granted. This results in a significant amount of human effort and time being wasted in the tender business procedures (Ng *et al.*, 2007).

Several e-tendering systems are well-known (Alyahya and Panuwatwanic, 2018). For example, Ng *et al.*'s (2007) model-tendering process uses the UML language. They also use an ad hoc diagram to describe the system's totality. They convert the UML class diagram into XML specifications for message exchange between stakeholders.

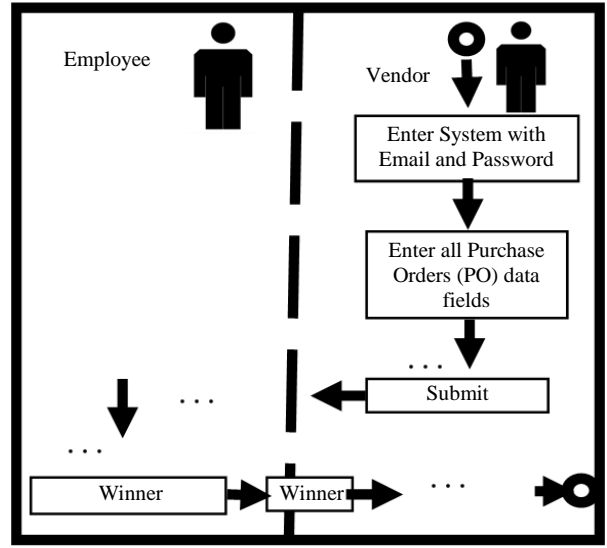

**Fig. 12:** Tender submission (redrawn, partially from Ng *et al.*, 2007)

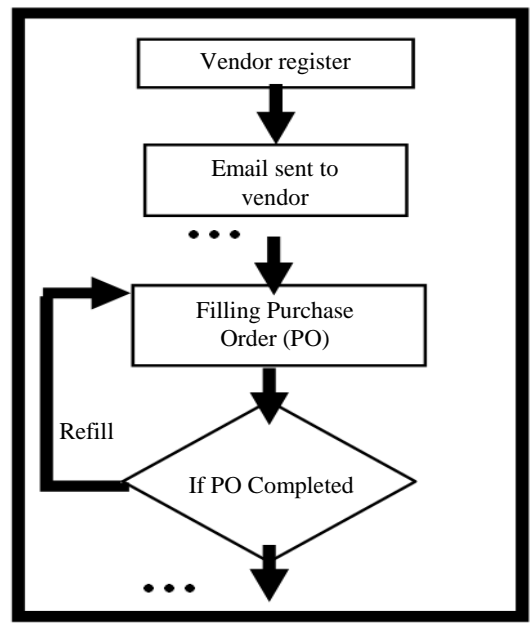

**Fig. 13:** Activity diagram for the tendering system (partially redrawn from (Alkhalifah and Ansari, 2016))

A general observation in the current tendering system model is the lack of a tool for building a holistic view of the system. According to Kong *et al.* (2009), "The intuitive nature of UML notations greatly facilitates distribution and communication of software artifacts among different developers." Although UML provides many notations, it is sufficient to use class diagrams and state diagrams (Cavarra *et al.*, 2003). Nevertheless, according to Cavarra *et al*. (2003), one weakness of UML is the lack of a well-defined model design process that integrates the different kinds of diagrams; it is "a bunch of notations without an effective integration." Furthermore, "This leads to a more apparent than real understanding of models, difficulty to perform rigorous analysis, validation, verification, integrity of models and difficulty to develop tools supporting mechanical validation and verification" (Cavarra *et al*., 2003).

Kong *et al*. (2009) proposed a visual approach that automatically assigns precise behavioral semantics to statechart diagrams. They defined an integrated behavior by combining the behavioral semantics of statechart diagrams with dynamic reconfigurations in object diagrams. The hierarchical structure of states is automatically formalized as a graph grammar (Kong *et al*., 2009).

In the TM approach, with its single diagram featuring events superimposed over a static description, integration is already a property of the system. Next, we show this feature for our case study. The following list includes some of the requirements specified by the stakeholder for the tendering system under consideration:

1. Vendor registration: Vendors must be able to electronically file all of their information and upload all of the required official documents
2. The tender data must include the vendor's name, date of submission, cost, legal requirements and other information
3. Tender openings must be provided
4. The Tendering Committee's meeting minutes must be provided
5. Rewarded tenders: A list of all tenders and the companies that they were rewarded to, with all related details, must be provided
6. Vendor qualifications must be available
7. Tender postponement must be an option
8. The e-tendering system must be ready for payment gateway integration, to allow purchases of tenders online
9. A minimum of ten reports on e-tendering in the intranet portal must be provided so that the tenders, vendors, categories, etc., can be reviewed
10. The system must provide a dashboard for different queries in the portal
11. The system must support an advanced workflow and fully utilize SharePoint technologies

According to the TM approach, the first task is to construct a grand representation of the system. Figure 14 shows the first part of the TM model, described as follows:

1. The vendor creates (Circle 1) a request for a registration account that flows to the system (2) to be processed after (3) creating an account (4) in the database. Then the account information is filled out with name, email, civil ID and password values and a description from the data supplied in the registration request (5).
2. Additionally, the following steps are triggered:

    (a) Initializing the account's status as *inactive* (6).
    (b) Initializing of the payment's fee-credit value as *zero* (7)

3. Then, an email value is created (8) that includes account data (name, civil ID, password and description) (9) and email destination (10)

The email flows to the vendor (11). Accordingly, the vendor is triggered (12) to go to the Nazaha building to pay its fees with a payment card (13). The card flows to the payment machine (14) to register the amount (15), which issues a receipt that flows (16) to the vendor. The receipt, then, is given to the NAZAHA employee (17), who inputs the payment number into the system (18). Upon the number's arrival in the system, the following processes are performed on the payment numbers:

i. The payment values are stored in the database (19)
ii. The payment value is inserted to the database by searching a table that contains all numbers of correspondence values
iii. The fee-credit value is changed to the *payment Value* (20)
iv. The user's status is switched to *active* account (21)
v. Then, the payment information is released to the system (22)

During the submission period (23), the vendor can request (24) to enter the system using its email address and password. The system will check if the vendor's account is activated and that it has paid its fees (25); then, a session will be created (26) for the vendor to enter the system (27). Meanwhile, the system involves the following:

(a) Initializing the application's status as *un-submitted* (28)
(b) Initializing the application's rank as *un-ranked* (29)
(c) Initializing the application's completion as *incomplete* (30)

The vendor will start by filling out the Purchase Order (PO) application (31) with all of the required fields (32) and send it back to the system (33) to be stored (34) and then view the full PO application (34) to submit it (35) and send it back to the system (36) to change the status to *submitted* (37).

**Fig. 14:** TM model of the case study tendering system

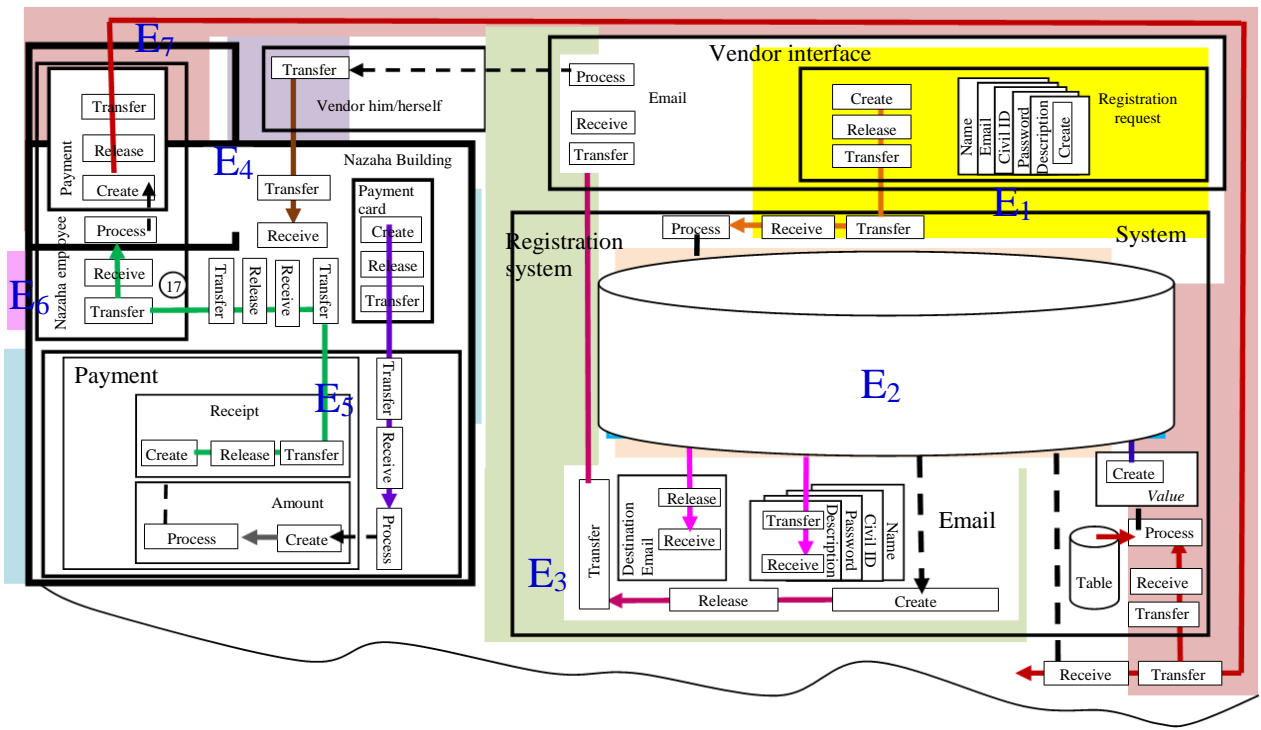

**Fig. 15:** Partial view of the events of the vendor-registration process in the tendering system

All PO applications are viewed (38) by the employee (39) for him or her to check whether they are *completed* (40) and then to rank them (41). The employee will view all of the ranked PO applications (42) to select the *winner* (43).

The static description of Fig. 14 can be converted into a program in any programming language (e.g., to C++, as described in Al-Fedaghi and Haidar, 2019). It is also used to identify the events.

For space considerations, Fig. 15 shows only the first seven events:

- Event 1 ($E_1$): The vendor requests registration in the tendering system
- Event 2 ($E_2$): The system creates a new account for the vendor
- Event 3 ($E_3$): An email is created and sent to the vendor
- Event 4 ($E_4$): The vendor receives the email and goes to the NAZAHA organization to complete its registration
- Event 5 ($E_5$): The vendor pays the registration fee and provides the receipt to the appropriate employee
- Event 6 ($E_6$): The employee receives the proof of payment and accesses the system
- Event 7 ($E_7$): The payment's serial number is inputted into the vendor's account to activate it

The resultant TM dynamic diagram can be used as a conceptual model in simulations, similar to using flowcharting in the simulation language Arena.

## Conclusion

We have applied the recently developed TM model as a conceptual framework to impose uniformity across the task of describing system requirements. The TM model, as a new type of philosophical foundation, incorporates complete unity between things and processes as well as five generic operations and integrates static and dynamic features of the system.

We introduced an enhanced version of the TM model and described its components and philosophical underpinnings.

We substantiated the model's viability using many examples from the literature and by modeling an actual case study. The TM model seems to provide new contributions to the field of conceptual modeling that can enhance and enrich current modeling methodologies such as OO and UML.

The complexity of the TM diagram may be considered a weakness of the approach. However, a TM diagram can be used at various levels of granularity and complexity, as in the case of nontechnical use. For example, Fig. 14 can be simplified by removing the release, transfer and receive stages under the assumption that the direction of the arrows is sufficient to represent the flow. This is demonstrated in the upper part of Fig. 16. Such simplification can be applied at various levels. The resulting diagram facilitates communication among various stakeholders and leads to a common understanding and mental picture of various system components.

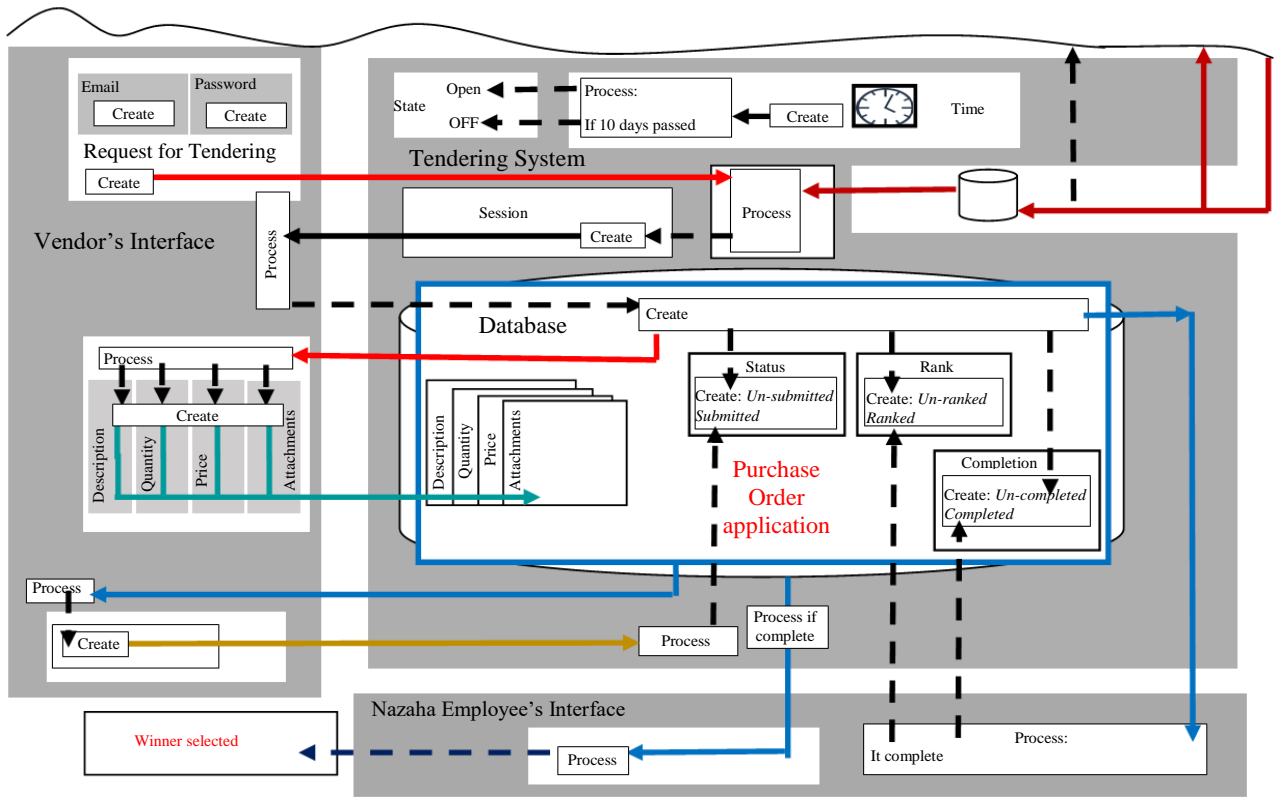

**Fig. 16:** Simplification of the lower part of the TM model of the case study tendering system

## Authors' Contributions

The first author is the main developer of the TM model. The second author is the main builder of the case study of the tendering system.

## Ethics

This article is original and contains unpublished material. The corresponding author confirms that all of the other authors have read and approved the manuscript. No ethical issues were involved and the authors have no conflict of interest to disclose.

## References

Al-Fedaghi, S., 2019a. Five generic processes for behaviour description in software engineering. Int. J. Comput. Sci. Inform. Security, 17: 120-131.

Al-Fedaghi, S., 2019b. Thing/machines (thimacs) applied to structural description in software engineering. Int. J. Comput. Sci. Inform. Security, 17: 1-11.

Al-Fedaghi, S., 2019c. Toward maximum grip process modeling in software engineering. Int. J. Comput. Sci. Inform. Security, 17: 8-18.

Al-Fedaghi, S. and G. Aldamkhi, 2019. Conceptual modeling of an IP phone communication system: A case study. Proceedings of the 18th Annual Wireless Telecommunications Symposium, Apr. 9-12, New York, NY, USA.

Al-Fedaghi, S. and O. Alsumait, 2019. Toward a conceptual foundation for physical security: Case study of an IT department. Int. J. Saf. Secur. Eng., 9: 137-156.

Al-Fedaghi, S. and J. Al-Fadhli, 2019. Modeling an unmanned aerial vehicle as a thinging machine. Proceedings of the 5th International Conference on Control, Automation and Robotics, Apr. 19-22, Beijing, China.

Al-Fedaghi, S. and M. BehBehani, 2018. Thinging machine applied to information leakage. Int. J. Adv. Comput. Sci. Applic., 9: 101-110.

Al-Fedaghi, S. and E. Haidar, 2019. Programming is diagramming is programming. J. Software, 14: 410-422.

Alkhalifah, A. and G.A. Ansari, 2016. Modeling of E-procurement System through UML using data mining technique for supplier performance. Proceedings of the International Conference on Software Networking, May 23-26, Jeju, South Korea, pp: 1-6.

Alyahya, M. and K. Panuwatwanic, 2018. Implementing e-tendering to improve the efficiency of public construction contract in Saudi Arabia. Int. J. Procurement Manage., 11: 267-267.

Bryant, L.R., 2012. Towards a machine-oriented aesthetics: On the power of art. The Matter of Contradiction Conference, Limousin, France.

Bunge, M.A., 1977. Treatise on Basic Philosophy: Ontology I-The Furniture of the World. 1st Edn., Kluwer Academic Publishers, Dordrecht, ISBN-10: 9027707855, pp: 354.

Cavarra, A., E. Riccobene and P. Scandurra, 2003. Integrating UML Static and Dynamic Views and Formalizing the Interaction Mechanism of UML State Machines. In: Abstract State Machines 2003 (ASM 2003), Börger, E., A. Gargantini and E. Riccobene (Eds.), Springer, Berlin, pp: 229-243.

Chen, P.P.S., 1976. The entity relationship model-toward a unified view of data. ACM Trans. Database Syst., 1: 9-36.

de Champeaux, D., D. Lea and P. Faure, 1993. Object-Oriented System Development. 1st Edn., Addison-Wesley, Reading, MA., ISBN-10: 020156355X, pp: 532.

DeLanda, M., 2015. The new materiality. Architectural Design, 85: 16-21.

DCS, 2010. Software engineering. University of Cape Town.

Eichelberger, H., 2003. Nice class diagrams admit good design? Proceedings of the ACM Symposium on Software Visualization, Jun. 11-13, ACM Press, San Diego, California, pp: 159-168.

Frank, U., 1999. Conceptual modelling as the core of the information systems discipline-perspectives and epistemological challenges. Proceedings of the 5th America's Conference on Information Systems, (CIS' 99), Association for Information Systems, Milwaukee, pp: 695-698.

Frigg, R. and J. Nguyen, 2017. Models and Representation. In: Springer Handbook of Model-Based Science, Magnani, L. and T. Bertolotti (Eds.), Berlin, Springer, pp: 49-102.

Grässle, P., H. Baumann and P. Baumann, 2005. UML 2.0 in action: A project-based tutorial. Packt Publishing Ltd., Birmingham, UK.

Guizzardi, G. and T.A. Halpin, 2008. Ontological foundations for conceptual modelling. Applied Ontol., 3: 1-12.

Heidegger, M., 1975. The Thing. In: Poetry, Language, Thought, Hofstadter, A. (Trans.), Harper and Row, New York, pp: 161-184.

Hölldobler, K., A. Roth, B. Rumpe and A. Wortmann, 2017. Advances in Modeling Language Engineering. In: Model and Data Engineering, Ouhammou, Y., M. Ivanovic, A. Abelló and L. Bellatreche (Eds.), Lecture Notes in Computer Science, pp: 3-17.

Hossain, B.A. and R. Schwitter, 2018. Specifying Conceptual Models Using Restricted Natural Language. In: Proceedings of Australasian Language Technology Association Workshop, Kim, S.M. and X. Zhang (Eds.), Dunedin, New Zealand, pp: 44-52.

Kong, J., K. Zhang, J. Dong and D. Xu, 2009. Specifying behavioral semantics of UML diagrams through graph transformations. J. Syst. Software, 82: 292-306.

Lauesen, S. and O. Vinter, 2001. Preventing requirement defects: An experiment in process improvement. Requirements Eng., 6: 37-50.

Lu, S. and J. Parsons, 2005. Enforcing ontological rules in UML-based conceptual modeling: Principles and implementation. Proceedings of the 10th Workshop on Evaluating Modeling Methods for Systems Analysis and Design, Held in Conjunction with the 17th Conference on Advanced Information Systems, (AIS' 05), FEUP, Porto, Portugal, pp: 451-462.

Meyer, N., 1997. Object-Oriented Software Construction. 2nd Ed. Prentice Hall Professional Technical Reference, ISBN-10: 0136291554.

Nath, R., 2020. Introduction to object-oriented methodology.

Ng, L.L.N., D.K.W. Chiu and P.C.K. Hung, 2007. Tendering Process Model (TPM) Implementation for B2B integration in a web services environment. Proceedings of the 40th Annual Hawaii International Conference on System Sciences, Jan. 3-6, Waikoloa, HI, USA, pp: 143-152.

OMG, 2000. Unified Modeling Language Specification. 1st Edn., OMG.

Osis, J. and E. Asnina, 2010. Model-Driven Domain Analysis and Software Development: Architectures and Functions. 1st Edn., IGI Global, Hershey, Pennsylvania, USA.

Recker, J.C., 2005. Conceptual model evaluation. Towards more paradigmatic rigor. Proceedings of the CAiSE Workshops, Jun. 13-14, Porto, Portugal.

Ribbert, M., B. Niehaves, A. Dreiling and R. Holten, 2004. An epistemological foundation of conceptual modeling. Proceedings of the 12th European Conference on Information Systems, Jun. 14-16, Turku, Finland, pp: 1-12.

Rumbaugh, J., M. Blaha, W. Premerlani, F. Eddy and W. Lorensen, 1991. Object-Oriented Modeling and Design. 3rd Edn., Prentice Hall, Englewood Cliffs, ISBN-10: 0136298419, pp: 500.

Singh, S.K., 2011. An event-based framework for object-oriented analysis, computation of metrics and identification of test scenarios. Ph.D. Thesis, Jaypee Institute of Information Technology, Deemed University, Noida, India.

Singh, S.K., S. Sabharwal and J.P. Gupta, 2010. An event-based methodology to generate class diagrams and its empirical evaluation. J. Comput. Sci., 6: 1301-1325.

Sfard, A., 1991. On the dual nature of mathematical conceptions: Reflections on processes and objects as different sides of the same coin. Educ. Stud. Math., 22: 1-36.

Steiner, H.G., 1952. Theory of mathematics education: An introduction. Learn. Math., 5: 11-17, 1985.

Stotts, D., 2007. Documenting an OO design: Class diagrams. team software engineering website.

Szlenk, M., 2006. Formal semantics and reasoning about UML class diagram. Proceedings of the International Conference on Dependability of Computer Systems, May 25-27, IEEE Xplore Press, Szklarska Poreba, Poland, pp: 51-59. DOI: 10.1109/DEPCOS-RELCOMEX.2006.27

Taivalsaari, A., 1997. Classes vs. prototypes: Some philosophical and historical observations. J. Object-Oriented Programm., 10: 44-50.

Wand, Y. and R. Weber, 2002. Research commentary: Information systems and conceptual modeling-a research agenda. Inform. Syst. Res., 13: 363-376.

Wand, Y. and R. Weber, 1993. On the ontological expressiveness of information systems analysis and design grammars. J. Inform. Syst., 3: 217-237. DOI: 10.1111/j.1365-2575.1993.tb00127.x

Washizaki, H., M. Akimoto, A. Hasebe, A. Kubo and Y. Fukazawa, 2010. TCD: A text-based UML class diagram notation and its model converters. Proceedings of the International Conference on Advanced Software Engineering and Its Applications, Dec. 13-15, Jeju Island, Korea, pp: 296-302.

Weisfeld, M., 2009. The Object-Oriented Thought Process. 3rd Edn., Addison-Wesley, Upper Saddle River, NJ.